\documentclass[12pt]{iopart}

\usepackage{cite}
\usepackage{iopams}
\usepackage{graphicx}
\expandafter\let\csname equation*\endcsname\relax
\expandafter\let\csname endequation*\endcsname\relax
\usepackage{amsmath}
\usepackage{array}
\usepackage{subfigure}
\usepackage{dcolumn}
\usepackage{bm}
\usepackage{sidecap}
\usepackage{CJK}
\usepackage{fancyhdr}

\begin{document}

%\title {A modified trusted homodyne detector model for continuous-variable quantum key distribution: detailed security analysis and improvement by phase-sensitive amplifier}
\title[A modified practical homodyne detector model for CV-QKD]{A modified practical homodyne detector model for continuous-variable quantum key distribution: detailed security analysis and improvement by the phase-sensitive amplifier}
%\title[\resizebox{4.5in}{!}{Long title}]{Long title}
%\author{Yundi Huang$^1$, Yichen Zhang$^{1}$, Luyu Huang$^{1}$ Song Yu$^{1}$}
\author{Yundi Huang, Yichen Zhang, Luyu Huang, Song Yu}
\address{State Key Laboratory of Information Photonics and Optical Communications, Beijing University of Posts and Telecommunications, Beijing 100876, China}

\ead{zhangyc@bupt.edu.cn}

\vspace{10pt}
\begin{indented}
\item[]May 2020
\end{indented}

\begin{abstract}
The practical homodyne detector model of continuous-variable quantum key distribution models the inherent imperfections of the practical homodyne detector, namely the limited detection efficiency and the electronic noise, into trusted loss. However, the conventional practical homodyne detector model is valid only when both the imperfections of the practical homodyne detector are calibrated.
In this paper, we show a modified practical homodyne detector model that can model the imperfections separately. The phase-sensitive amplifier is further applied to compensate the imperfections of the practical homodyne detector. The feasibility of the modified practical homodyne detector model with the phase-sensitive amplifier is proved and the security analysis is provided in detail. Simulation results reveal that the phase-sensitive amplifier can be used to improve the performance of the modified practical homodyne detector model, and when the gain is infinitely high, the limited detection efficiency can be fully compensated.

\end{abstract}

%\begin{multicols}{2}
\section{Introduction}
Quantum key distribution (QKD)~\cite{pirandola2019advances, xu2019secure} which aims at establishing secure key distribution process is one of the most practical applications in the field of quantum information science.
Continuous variable (CV) QKD~\cite{weedbrook2012gaussian, diamanti2015distributing} which is developed slightly posterior to discrete variable QKD, is now going through a booming period.
Many kinds of CV-QKD protocols with specific purposes are proposed and analyzed: measurement-device-independent CV-QKD protocols can defense arbitrary attacks against the detector~\cite{li2014continuous, zhang2014continuous, pirandola2015high}, while source-device-independent CV-QKD protocols are intrinsically secure against the malicious source~\cite{weedbrook2012continuous, zhang2020continous}. Discrete modulation CV-QKD protocols can effectively reduce the modulation complexity~\cite{li2018user, ghorai2019asymptotic, lin2019asymptotic}, and passive state preparation CV-QKD protocols use the thermal source to generate quantum states which can reduce the difficulty of the state preparation~\cite{qi2018passive, qi2020experimental}. Studies have also been extended to the attack scheme of the eavesdropper Eve~\cite{tserkis2020teleportation} and the future QKD network model~\cite{li2020mathematical}.
The coherent states and homodyne detection CV-QKD (GG02)~\cite{grosshans2002continuous, grosshans2003quantum} protocol is currently the most applicable protocol in practical implementations. It can apply off-the-shelf components to prepare coherent states and detect coherent states with the high bandwidth homodyne detector.
Experimental demonstration of over 200km of transmission distance under the laboratory conditions has been reported~\cite{zhang2020long}. The practical field test has reached 50km which proves the applicability of the coherent states and homodyne detection CV-QKD protocol in a metropolitan distance~\cite{zhang2019continuous}.

The practical implementations of the CV-QKD protocols will inevitably encounter the gap between the theoretical entanglement-based (EB) model and the practical prepare-and-measure (PM) model. The practical implementations suffer from different kinds of imperfections of the actual devices. These imperfections may threat the practical security of the CV-QKD system and degrade the secret key rate~\cite{xu2019secure}. Previously, the security analysis assumes any loss and noise are attributed to the eavesdropping in the quantum channel, which results in a very low secret key rate. Great efforts have been dedicated to fulfil this gap, and currently the most applied approach is to characterize the practical devices so that the imperfections can be properly modelled in the theoretical EB model~\cite{usenko2016trusted}.
Since the devices are normally within the control of the trusted parties, by carefully characterizing and calibrating the devices, we are able to model the imperfections as trusted noise and loss in the EB model. In this way, the negative impacts of the imperfections can be mitigated compared to the previous analysis that treats the imperfections as untrusted channel loss.
Many studies have been conducted to characterize the imperfections of the practical devices and model these imperfections in the EB model. For the sender Alice, noise introduced during the practical state preparation has been studied~\cite{filip2008continuous, shen2009security, usenko2010feasibility, shen2011continuous}, the corresponding monitor scheme is also proposed~\cite{yang2012source}. For the receiver Bob, the major imperfections of the practical homodyne detector are the limited detection efficiency and the electronic noise, which have been analyzed and modelled as trusted detection noise and trusted detection loss in the EB model~\cite{lodewyck2007quantum, usenko2016trusted}.
%By modelling the characterized practical devices, the performance of the practical CV-QKD protocols can be improved compared to the case where the imperfections are treated as untrusted channel losses.

However, the conventional practical homodyne detector model works only when both the detection efficiency and the electronic noise are calibrated. It means that even though the detection efficiency is calibrated, it still cannot be treated as trusted loss if the electronic noise is not properly calibrated. In this paper, we show a modified practical homodyne detector model that models the detection efficiency and the electronic noise separately. This model exhibits great importance because it allows any calibrated imperfections to be modelled as trusted loss, which makes the practical homodyne detector model more flexible. The relation between the modified practical homodyne detector model and the conventional practical homodyne detector model is studied in detail, while we show that the modified practical homodyne detector model can describe the practical PM model. The one-time-calibration method~\cite{zhang2020one} can be viewed as a particular instance that directly benefits from the modified practical homodyne detector model.
We further improve the modified practical homodyne detector model by applying the phase-sensitive amplifier (PSA).
Several studies has suggested that the conventional practical homodyne detector model can be improved by the PSA in different scenarios~\cite{fossier2009improvement, zhang2014improvement, wang2019continuous, huang2019improvement}.
Subsequently, it is natural for us to wonder whether such an amplifier can be used in the modified practical homodyne detector model. We thoroughly analysis the possibility of applying the PSA to increase the secret key rate of the CV-QKD protocol that is based on the modified practical homodyne detector model. The feasibility and detailed secret key rate analysis are provided, numerous simulation results are demonstrated to show the effect of the PSA.

The rest of the paper is organised as follows. In section 2, the modified practical homodyne detector model is introduced, and the relation with the conventional practical homodyne detector model is addressed. In section 3, the PSA is applied to the modified practical homodyne detector model. We prove the feasibility and provide detailed secret key rate analysis. Numerous simulation results are shown in section 4 where the performance of the modified practical homodyne detector model with the assistance of the PSA is analyzied. Conclusions are drawn in section 5.

\section{The modified practical homodyne detector model}
In this section, the modified practical homodyne detector model is introduced right after a short review on the conventional practical homodyne detector model. By exploiting the relation of the electronic noise between them, we prove that the two practical homodyne detector models are essentially the same, when the raw data from the practical realization is normalized by the same shot-noise unit (SNU). Lastly, we discuss how to model single imperfection of the practical homodyne detector individually by applying the modified practical homodyne detector model.
We restrict ourselves to the coherent states and homodyne detection scheme, where the receiver Bob randomly measures ${x}$ or ${p}$ quadrature of the signal. Two major imperfections of a practical homodyne detector are considered: the limited detection efficiency and the electronic noise.

\begin{figure}
\centerline{\includegraphics[width=16cm]{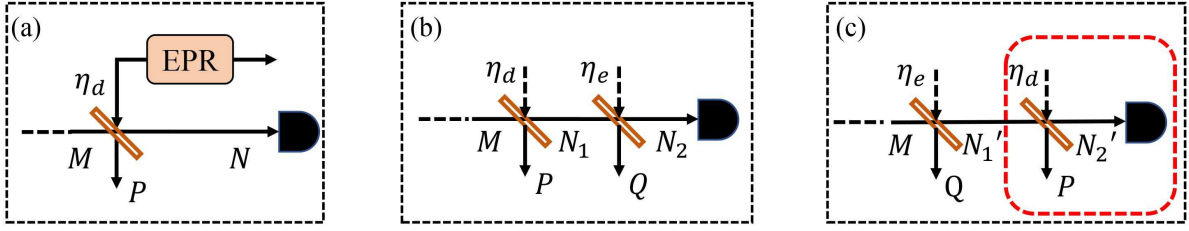}}
\caption{Three ways of modelling the practical homodyne detector. (a) The conventional practical homodyne detector model. The limited detection efficiency is imitated by the transmittance of the BS, while an EPR state is applied to model the electronic noise, and one of its mode is injected to the other port of the BS, the introduced noise is used to imitate the electronic noise. (b) The modified practical homodyne detector model. The first BS with the transmittance ${\eta_d}$ still imitates the detection efficiency, while the electronic noise is modelled by another BS with the transmittance ${\eta_e}$. (c) A variation of the modified practical homodyne detector model. In this scenario, the incoming signal mode first passes through the BS that represents the electronic noise, then passes through the BS that represents the detection efficiency.
}
\end{figure}

The PM model is normally applied in the practical implementations, while the EB model is used to perform security analysis and secret key rate calculation. The security relies on the equivalence between the PM model and the EB model. Yet, the raw data obtained from the PM model cannot be taken into the corresponding EB model directly. It has to be normalized by the SNU first.  More precisely, the output variance of the PM model after the SNU normalization should equal to that of the EB model, so that the equivalence can hold.
The EB model of the conventional practical homodyne detector model is depicted in figure 1(a), the limited detection efficiency ${\eta_d}$ is modelled by the transmittance of the beamsplitter (BS) while the electronic noise is modelled by an Einstein{-}Podolsky{-}Rosen (EPR) state whose one mode is injected to the other port of the BS, the introduced noise variance in mode ${N}$ is set as the electronic noise ${v_{el}}$.
It can be easily calculated that the variance of the output from this EB model is
\begin{equation}
V_{{N}}^{EB} = {\eta _d}{V_{{M}}} + (1 - {\eta _d}) + v{}_{el},
\end{equation}
where the variance of the EPR state is set as ${1 + \frac{{v{}_{el}}}{{1 - {\eta _d}}}}$. The equivalence with the PM model has been analyzed in ~\cite{lodewyck2007quantum, fossier2009improvement}.

We now turn to the modified practical homodyne detector model which is shown in figure 1(b). In this model, both the detection efficiency and the electronic noise are imitated by the BSs. The incoming signal ${M}$ first passes through the BS with the transmittance of ${\eta_d}$ then passes through the BS with the transmittance of ${\eta_e}$. In the following, we study the relation between the modified practical homodyne detector model and the conventional practical homodyne detector model while showing that the modified practical homodyne detector model can describe the practical PM model. It can be observed that both the conventional and the modified practical homodyne detector model use the transmittance of the BS to imitate the detection efficiency of the practical homodyne detector, the conventional model uses an EPR state to imitate the electronic noise while the modified model uses the transmittance of another BS to imitate the electronic noise. Thus, we focus on the relation of the electronic noise of the two models. As is analyzed in ~\cite{zhang2020one}, the electronic noise ${\eta_e}$ in the modified practical homodyne detector model is defined as
\begin{equation}
{\eta _e} = \frac{{{A^2}X_{LO}^2}}{{{A^2}X_{LO}^2 + {V_{ele}}}},
\end{equation}
where the parameters ${A}$, ${X_{LO}}$ and ${V_{ele}}$ are all corresponding to the practical realizations in the PM model. The parameter ${A}$ describes the amplification of the practical homodyne detector, ${X_{LO}}$ represents the local oscillator (LO) that interferences with the quantum signal, and ${V_{ele}}$ is the variance of the raw electronic noise of the practical homodyne detector. However, this equation does not directly reflect the relation of the electronic noise appeared in the conventional practical homodyne detector model and the modified practical homodyne detector model. The electronic noise ${v_{el}}$ used in the conventional practical homodyne detector model is the value of the raw electronic noise ${V_{ele}}$ normalized by the SNU. The SNU in the conventional practical homodyne detector model is defined as ${{u_s} = {A^2}X_{LO}^2}$. By dividing ${{A^2}X_{LO}^2}$ on both the numerator and the denominator in equation (2), we obtain the relation between ${\eta_e}$ and ${v_{el}}$
\begin{equation}
{\eta _e} = \frac{1}{{1 + {v_{el}}}}.
\end{equation}
Now, we have obtained the relation of the electronic noise defined by the two models. The electronic noise from the conventional practical homodyne detector model can be directly transformed into the modified practical homodyne detector model, characterized by the transmittance of the BS through equation (3). Yet, the SNU used in the conventional practical homodyne detector model and that used in the modified practical homodyne detector model are not the same. In the modified practical homodyne detector model, the SNU is defined as ${{u_s}^\prime  = {A^2}X_{LO}^2 + {V_{ele}}}$. In order to describe the output of the practical homodyne detector after being normalized by the conventional SNU ${{u_s}}$, we define a scaling parameter ${s = {u_s}^\prime /{u_s} = 1 + {v_{el}}}$ that represents the transformation from the modified SNU ${{u_s}^\prime}$ to the conventional SNU ${{u_s}}$. Now we consider the output of the modified practical homodyne detector model
\begin{equation}
V_{{N_2}}^{EB} = {\eta _e}{\eta _d}{V_{{M}}} - {\eta _e}{\eta _d} + 1.
\end{equation}
By taking equation (3) into equation (4), then multiplying the scaling parameter ${s}$, it can be easily seen that the outcomes are exactly the same for the two models. So, we conclude that even though the modified practical homodyne detector model corresponds to a different SNU ${{u_s}^\prime}$, it can still describe the practical homodyne detector that is normalized by the conventional SNU ${{u_s}}$. In other words, the two models are essentially the same given the fact that the output of the practical homodyne detector is normalized by the same SNU.

It should be pointed out that the modified practical homodyne detector model can be directly applied as long as the raw data from the PM model is normalized by the SNU ${{u_s}^\prime}$. As can be seen from the EB model, the modified practical homodyne detector model models the imperfections of the practical homodyne detector separately.
The imperfection of the electronic noise can be properly modelled even if the detection efficiency is not characterized. When the detection efficiency is not characterized, it is treated as untrusted loss, while the electronic noise can still be modelled as trusted loss in the modified practical homodyne detector model.

We can further transform the modified practical homodyne detector model by swapping the order of the BSs. In this EB model, the incoming quantum signal first passes through the BS that represents the electronic noise, then passes through the BS that represents the detection efficiency, as is depicted in figure 1(c).
Based on the basic assumptions made on the practical homodyne detector model, this operation will not disturb the security of the system. One can easily check that the output variance of this model is exactly the same as the model in figure 1(b).
Thus the EB model after the swapping operation can also be applied to model the imperfections of the practical homodyne detector. This model allows us to model the detection efficiency as trusted loss individually if the electronic noise is not properly characterized. It means that we can perform the security analysis without knowing the exact value of the electronic noise, and maximally retain the advantages of using the practical homodyne detector model.
Moreover, this model offers extra advantages: it will no longer be required to measure the electronic noise to calibrate the SNU in the corresponding PM model. The accuracy of the calibrated SNU can be improved, and the spectral efficiency is increased~\cite{zhang2020one}. Therefore, the following analysis will focus on this particular EB model.

\section{Detailed security analysis and improvement by the phase-sensitive amplifier}
In this section, we improve the performance of the modified practical homodyne detector model by using the PSA. In subsection 3.1,
the EB model of the modified practical homodyne detector model with the PSA is proposed and proved to be equivalent with the practical implementations. Detailed secret key rate analysis is provided in subsection 3.2.

\subsection{Security proof of applying the PSA with the modified practical homodyne detector model}
In this subsection we first review the principle of the PSA then apply it to the modified practical homodyne detector model.
The PSA~\cite{caves1982quantum, tong2011towards} that has been extensively studied can ideally amplify the target canonical quadrature while squeezing the other canonical quadrature. In this paper, we assume that ${x}$ quadrature is the target quadrature to be amplified, the amplification process can be described as ${\hat x \to \sqrt g \hat x;\;\hat p \to \sqrt {1/g} \hat p}$, where the parameter ${g}$ stands for the gain factor, for any value that ${g \ge 1}$, the target quadrature is amplified.
The practical implementations of the PSA may introduce other noises in the process, yet, these effects are neglected in this paper.

\begin{figure}
\centerline{\includegraphics[width=14cm]{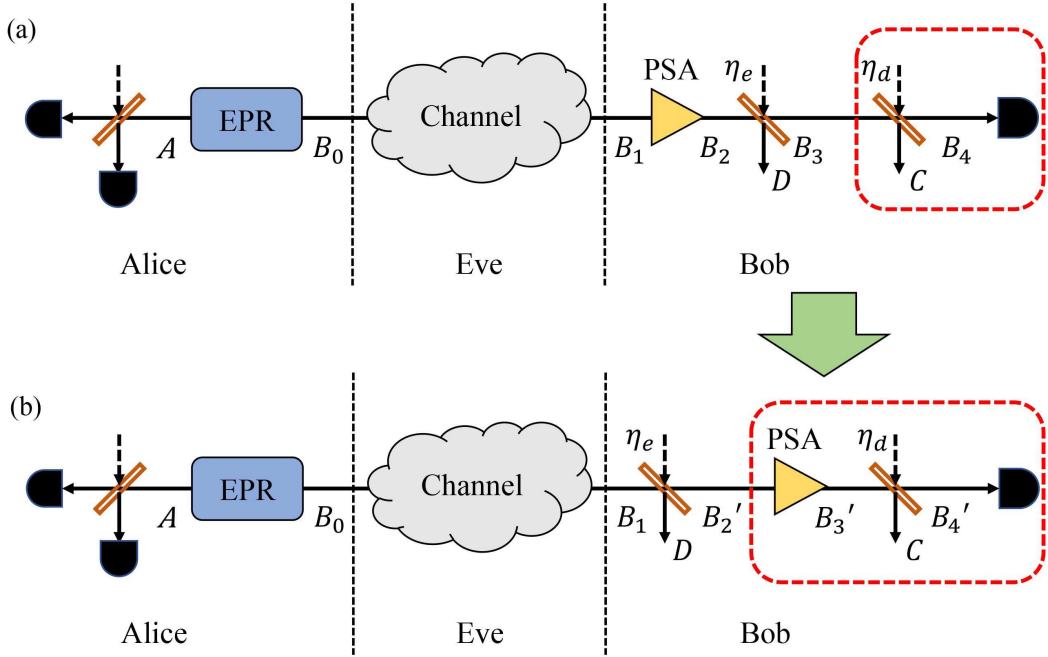}}
\caption{Complete EB model of the modified practical homodyne detector model with the PSA. Alice generates an EPR state then heterodyne detects one of its mode, which will project the other mode into a coherent state. Alice sends the coherent state to Bob. (a) The PSA is placed before the modified practical homodyne detector model. The incoming signal is first amplified by the PSA then in turn passes through the BS that represents the electronic noise and the BS that represents the detection efficiency before being detected by the ideal homodyne detector. (b) The PSA is placed between the BS that represents the electronic noise and the BS that represents the detection efficiency. The incoming signal first passes through the BS that represents the electronic noise then gets amplified by the PSA. Next, the signal passes through the BS that represents the detection efficiency before being detected by the ideal homodyne detector.
}
\end{figure}

In practical implementations, the PSA is placed before the practical homodyne detector.
Since the PSA is placed right before the practical homodyne detector in the PM model, in its corresponding EB model, one would expect that the PSA should also be placed right before practical homodyne detector model, as is shown in figure 2(a). It is obvious that this EB model is equivalent to the PM model.
Yet, the electronic noise ${\eta_e}$ is unknown in the PM model, which means the mode ${D}$ in the EB model is also unknown, thus anything before the BS that represents the electronic noise cannot be directly taken into the secret key rate calculation. Thus, the PSA in this scheme is not trusted in the security analysis.
Following the analysis in section 2, we swap the order of the PSA and place it in the middle of the two BSs, as is depicted in figure 2(b), where the PSA is placed in the middle of the BS that represents the electronic noise and the BS that represents the detection efficiency.
The feasibility of swapping the order of the PSA and the BS is based on the followings: the swapping operation will not affect the output of the EB model, which means that equivalence between the EB model and PM model still holds. The swapping operation is taken place inside Bob, which means that the operation cannot affect the amount of information that Eve can obtain.
After the swapping operation, the PSA can be considered trusted in the secret key rate calculation.
One may concern whether this EB model is identical to the practical implementations of the PM model, in the following, we provide the detailed proof.

We start with the PM model, the incoming signal is first compensated using the PSA then gets detected by the practical homodyne detector. The outputs of the practical homodyne detector after the PSA are
\begin{equation}
\begin{array}{l}
x_{out}^{PM} = A{X_{LO}}(\sqrt {{\eta _d}} \sqrt g {\hat x_{{B_1}}} + \sqrt {1 - {\eta _d}} {{\hat x}_{v_1}}) + {X_{ele}},\\
p_{out}^{PM} = A{X_{LO}}(\sqrt {{\eta _d}} \sqrt {1/g} {\hat p_{{B_1}}} + \sqrt {1 - {\eta _d}} {{\hat p}_{v_1}}) + {P_{ele}},
\end{array}
\end{equation}
where ${\hat x_{{B_1}}}$ and ${\hat p_{{B_1}}}$ are the canonical quadratures of mode ${B_1}$, ${{{\hat x}_{v_1}}}$ and ${{{\hat p}_{v_1}}}$ are the corresponding quadratures of the vacuum, ${X_{ele}}$ and ${P_{ele}}$ represent the practical electronic noise in ${x}$ quadrature and ${p}$ quadrature respectively. Both the vacuum and electronic noise are Gaussian variables with symmetrical behaviours on both quadratures, the variance of the vacuum is 1, the variance of the raw electronic noise is ${V_{ele}}$.
The raw data from the PM model needs to be normalized by the SNU ${{u_s}^\prime}$, so that the outputs after the normalization would be
\begin{equation}
\begin{array}{l}
x_{{u_s}^\prime }^{PM} = \frac{{A{X_{LO}}}}{{\sqrt {{A^2}X_{LO}^2 + {V_{ele}}} }}(\sqrt {{\eta _d}} \sqrt g {\hat x_{{B_1}}} + \sqrt {1 - {\eta _d}} {{\hat x}_{v_1}}) + \frac{{{X_{ele}}}}{{\sqrt {{A^2}X_{LO}^2 + {V_{ele}}} }},\\
p_{{u_s}^\prime }^{PM} = \frac{{A{X_{LO}}}}{{\sqrt {{A^2}X_{LO}^2 + {V_{ele}}} }}(\sqrt {{\eta _d}} \sqrt {1/g} {\hat p_{{B_1}}} + \sqrt {1 - {\eta _d}} {{\hat p}_{v_1}}) + \frac{{{P_{ele}}}}{{\sqrt {{A^2}X_{LO}^2 + {V_{ele}}} }}.
\end{array}
\end{equation}
The Gaussian variables ${X_{ele}}$ and ${P_{ele}}$ can be further rewrite as ${\sqrt {{V_{ele}}} {\hat x_{v_2}}}$ and ${\sqrt {{V_{ele}}} {\hat p_{v_2}}}$, where ${{\hat x_{v_2}}}$ and ${{\hat p_{v_2}}}$ also represent the canonical quadratures of the vacuum. Then by substituting ${\frac{{A{X_{LO}}}}{{\sqrt {{A^2}X_{LO}^2 + {V_{ele}}} }}}$ with ${\sqrt {{\eta _e}}}$, as in equation (2), we may write out the final outputs with regard to the BSs with ${\eta_e}$ and ${\eta_d}$
\begin{equation}
\begin{array}{l}
x_{{u_s}^\prime }^{PM} = \sqrt {{\eta _e}} (\sqrt {{\eta _d}} \sqrt g {\hat x_{{B_1}}} + \sqrt {1 - {\eta _d}} {{\hat x}_{v_1}}) + \sqrt {1 - {\eta _e}} {{\hat x}_{v_2}},\\
p_{{u_s}^\prime }^{PM}= \sqrt {{\eta _e}} (\sqrt {{\eta _d}} \sqrt {1/g} {\hat p_{{B_1}}} + \sqrt {1 - {\eta _d}} {{\hat p}_{v_1}}) + \sqrt {1 - {\eta _e}} {{\hat p}_{v_2}}.
\end{array}
\end{equation}
Next, we consider the corresponding EB model. After the swapping operation, the incoming signal ${B_1}$ first passes through the BS with the transmittance ${\eta_e}$, then gets compensated by the PSA, finally passes through the BS that imitates the detection efficiency ${\eta_d}$. The outputs for ${x}$ quadrature and ${p}$ quadrature would be
\begin{equation}
\begin{array}{l}
x_{out}^{EB} = \sqrt {{\eta _e}} (\sqrt {{\eta _d}} \sqrt g {\hat x_{{B_1}}} + \sqrt {1 - {\eta _d}} {{\hat x}_{v_1}}) + \sqrt {1 - {\eta _e}} {{\hat x}_{v_2}},\\
p_{out}^{EB}= \sqrt {{\eta _e}} (\sqrt {{\eta _d}} \sqrt {1/g} {\hat p_{{B_1}}} + \sqrt {1 - {\eta _d}} {{\hat p}_{v_1}}) + \sqrt {1 - {\eta _e}} {{\hat p}_{v_2}}.
\end{array}
\end{equation}
Since the security analysis requires the second order statistics, by calculating the variance of the output, it can be verified that the outputs of the PM model and the EB model have the same variance on the detected mode. The equivalence between the PM model and the EB model still holds. Now, we have proven the feasibility that the EB model in figure 2(b) can be used to characterize the PM model.

%之前还有一块儿没写，就是说可以交换那里；
We specifically place the PSA in the middle of the two BSs in the EB model, not directly place it at the output of the quantum channel like in the corresponding PM model. This is because the PSA can only compensate the trusted loss in the practical homodyne detector. Since in this model, only the the detection efficiency is modelled as trusted loss, the PSA is placed right before the BS that represents the detection efficiency to compensate the limited detection efficiency.

\subsection{Secret key rate analysis}

The central idea of conducting security analysis from the EB model is to estimate the secret key rate of the corresponding PM model. The secret key rate can be calculated as~\cite{devetak2005distillation}
\begin{equation}
R = \beta {I_{AB}} - {\chi _{BE}},
\end{equation}
where ${\beta}$ is the reconciliation efficiency, ${I_{AB}}$ is the classical mutual information between the two legitimate parties while ${\chi_{BE}}$ describes the quantum Von Neumann entropy that Eve gains in the reverse reconciliation scheme.

${I_{AB}}$ can be calculated from the variance and conditional variance of Alice and Bob
\begin{equation}
{I_{AB}}=\frac{{\rm{1}}}{{\rm{2}}}\log \frac{{{V_{{A^M}}}}}{{{V_{{A^M}|{B_4}^\prime }}}},
\end{equation}
where ${A^M}$ represents the mode A after being heterodyne detected, and ${{{V_{{A^M}}}}}$ is its variance. ${{{V_{{A^M}|{B_4}^\prime }}}}$ is the conditional variance of ${{A^M}}$ after the measurement of mode ${{B_4}^\prime}$, and it can be calculated as ${{V_{{A^M}|{B_4}^\prime }} = {V_{{A^M}}} - \frac{{ < {A^M}{B_4}^\prime  > }}{{2{V_{{B_4}^\prime }}}}}$, where ${{ < {A^M}{B_4}^\prime  > }}$ is the co-variance of ${{{A^M}}}$ and ${{B_4}^\prime}$, and all of the variances and co-variances can be found in the corresponding co-variance matrix ${{\gamma _{AC{B_4}^\prime}}}$.

Next, we calculate ${\chi _{BE}}$ while providing the detailed derivation of the co-variance matrix ${{\gamma _{AC{B_4}^\prime}}}$. ${\chi _{BE}}$ can be calculated from the following
\begin{equation}
{\chi _{BE}} = S(E) - S(E|x_{{{B_4}}^\prime}^M),
\end{equation}
where ${S( \cdot )}$ stands for the Von Neumann entropy which can be calculated from the symplectic eigenvalues of the corresponding co-variance matrix. So, ${\chi _{BE}}$ can also be determined from
\begin{equation}
{\chi _{BE}} = \sum\nolimits_{i = 1}^3 G (\frac{{{\lambda _i} - 1}}{2}) - \sum\nolimits_{i = 4}^5 G (\frac{{{\lambda _i} - 1}}{2}).
\end{equation}
The function ${G(x) = (x + 1){\log _2}(x + 1) - x{\log _2}x}$, and ${{\lambda _i}}$ are the symplectic eigenvalues of the corresponding co-variance matrix.
In the process of estimating the amount of entropy that Eve gains, we assume that Eve has unlimited power which gives her the ability to purify the entire system. Thus, one may write ${S(E) = S(ADC{{B_4}}^\prime)}$. What is being tricky is that since in the PM model, the electronic noise is not measured, the $\eta_e$ and the mode ${D}$ in the EB model are unknown to us. Therefore, the purification of Eve should be rewritten as ${S(E) = S(AC{B_4}^\prime )}$, which can be deduced from the co-variance matrix ${{\gamma _{AC{B_4}^\prime}}}$. In order to finally obtain the co-variance matrix ${{\gamma _{AC{B_4}^\prime}}}$, we start from the co-variance ${{\gamma _{A{B_1}}}}$, where ${B_1}$ represents the quantum signal that just passes through the quantum channel
\begin{equation}
{\gamma _{A{B_1}}} = \left( {\begin{array}{*{20}{c}}
{V\mathbb{I}}&{\sqrt {T({V^2} - 1)} {\sigma _z}}\\
{\sqrt {T({V^2} - 1)} {\sigma _z}}&{[T(V - 1 + {\varepsilon _c}) + 1]\mathbb{I}}
\end{array}} \right),
\end{equation}
the parameter ${V}$ is the variance of the initial EPR state, the channel parameters are the channel transmittance ${T}$ and the channel excess noise ${{\varepsilon _c}}$, ${\mathbb{I}}$ is the 2*2 identity matrix and ${{{\sigma _z}}}$ is the Pauli ${z}$ operator.
The mode ${B_1}$ then passes through the first BS with the transmittance ${\eta_e}$, which can be derived as
\begin{equation}
{\gamma _{A{{B_2}}^\prime D}} = Y_{{\eta _e}}^{BS}*[{\gamma _{A{B_1}}} \otimes \mathbb{I}]*{(Y_{{\eta _e}}^{BS})^T},
\end{equation}
where ${Y_{{\eta _e}}^{BS}}$ describes the operation of the BS on mode ${{B_1}}$ and the introduced vacuum, ${Y_{{\eta _e}}^{BS}}$ can be derived as: ${Y_{{\eta _e}}^{BS} = {\mathbb{I}} \otimes {Y_{{\eta _e}}}}$.
The matrix ${{Y_{{\eta _e}}}}$ represents the transformation of the first BS ${{Y_{{\eta _e}}} = \left( {\begin{array}{*{20}{c}}
{\sqrt {{\eta _e}} \mathbb{I}}&{\sqrt {1 - {\eta _e}} \mathbb{I}}\\
{ - \sqrt {1 - {\eta _e}} \mathbb{I}}&{\sqrt {{\eta _e}} \mathbb{I}}
\end{array}} \right)}$.
However, only the co-variance matrix ${{\gamma _{A{{B_2}}^\prime}}}$ can be obtained in practice because the mode ${D}$ is unknown to us. The mode ${{{B_2}}^\prime}$ is then compensated by the PSA. Since we have assumed the homodyne detector always measures the ${x}$ quadrature, the transformation of the PSA on the quadrature level can be described as
\begin{equation}
{Y^{PSA}} = \left( {\begin{array}{*{20}{c}}
{\sqrt g }&0\\
0&{1/\sqrt g }
\end{array}} \right).
\end{equation}
The co-variance matrix ${{\gamma _{A{{B_3}}^\prime}}}$ can be obtained after mode ${{B_2}^\prime}$ is compensated by the PSA
\begin{equation}
{\gamma _{A{{B_3}}^\prime}} = Y_g^{PSA}*{\gamma _{A{{B_2}}^\prime}}*{(Y_g^{PSA})^T},
\end{equation}
where the transformation matrix ${Y_g^{PSA}}$ can be derived as ${Y_g^{PSA} = {\mathbb{I}} \otimes {Y^{PSA}}}$.
Next, the mode ${{B_3}^\prime}$ passes through the last BS whose transmittance is the detection efficiency ${\eta_d}$. The co-variance matrix ${{\gamma _{A{{B_4}}^\prime C}}}$ is calculated as
\begin{equation}
{\gamma _{A{{B_4}}^\prime C}} = Y_{{\eta _d}}^{BS}*[{\gamma _{A{{B_3}}^\prime}} \otimes \mathbb{I}]*{(Y_{{\eta _d}}^{BS})^T},
\end{equation}
where ${Y_{{\eta _d}}^{BS}}$ can be derived from ${Y_{{\eta _d}}^{BS} = {\mathbb{I}} \otimes {Y_{{\eta _d}}}}$, it means that the BS is acting on mode ${{B_3}^\prime}$ and the vacuum. The transformation of the last BS is ${{Y_{{\eta _d}}} = \left( {\begin{array}{*{20}{c}}
{\sqrt {{\eta _d}} \mathbb{I}}&{\sqrt {1 - {\eta _d}} \mathbb{I}}\\
{ - \sqrt {1 - {\eta _d}} \mathbb{I}}&{\sqrt {{\eta _d}} \mathbb{I}}.
\end{array}} \right)}$.
Finally, we can obtain the co-variance matrix ${{\gamma _{AC{{B_4}}^\prime}}}$ by rearranging the order of the elements in ${{\gamma _{A{{B_4}}^\prime C}}}$. Thus, the first part of the ${{\chi _{BE}}}$ can be figured out by calculating the symplectic eigenvalues of ${{\gamma _{AC{{B_4}}^\prime}}}$.
The second term of ${{\chi _{BE}}}$ is the remaining Von Neumann entropy of Eve given the fact that the mode ${{B_4}^\prime}$ is being homodyne detected.
It can be deduced from the co-variance matrix $\gamma _{AC}^{{m_{{{B_4}^\prime}}}}$ which is derived as
\begin{equation}
\gamma _{AC}^{{m_{{{B_4}}^\prime}}} = {\gamma _{AC}} - {\sigma _{AC}}{(X{\gamma _{{{B_4}}^\prime}}X)^{MP}}\sigma _{AC}^T,
\end{equation}
${X = {\rm{diag}}(1,\;0,\;1,\;0)}$ and ${MP}$ denotes the inverse on the range. The matrix ${{\gamma _{AC}}}$, ${{\sigma _{AC}}}$ and ${\gamma_{{B_4}^\prime}}$ are the partitions of the co-variance matrix ${{\gamma _{AC{{B_4}}^\prime}}}$
\begin{equation}
{\gamma _{AC{{B_4}}^\prime}} = \left( {\begin{array}{*{20}{c}}
{{\gamma _{AC}}}&{{\sigma _{AC}}}\\
{\sigma _{AC}^T}&{{\gamma _{{{B_4}}^\prime}}}
\end{array}} \right).
\end{equation}

By calculating the eigenvalues of ${\gamma _{AC}^{{m_{{{B_4}}^\prime}}}}$, and inserting them into equation (12), we are able to calculate ${S(E|x_{{{B_4}}^\prime}^M)}$. Therefore, the complete secret key rate can be determined.

\section{Simulation results and discussion}
\begin{figure}[t]
\centerline{\includegraphics[width=10cm]{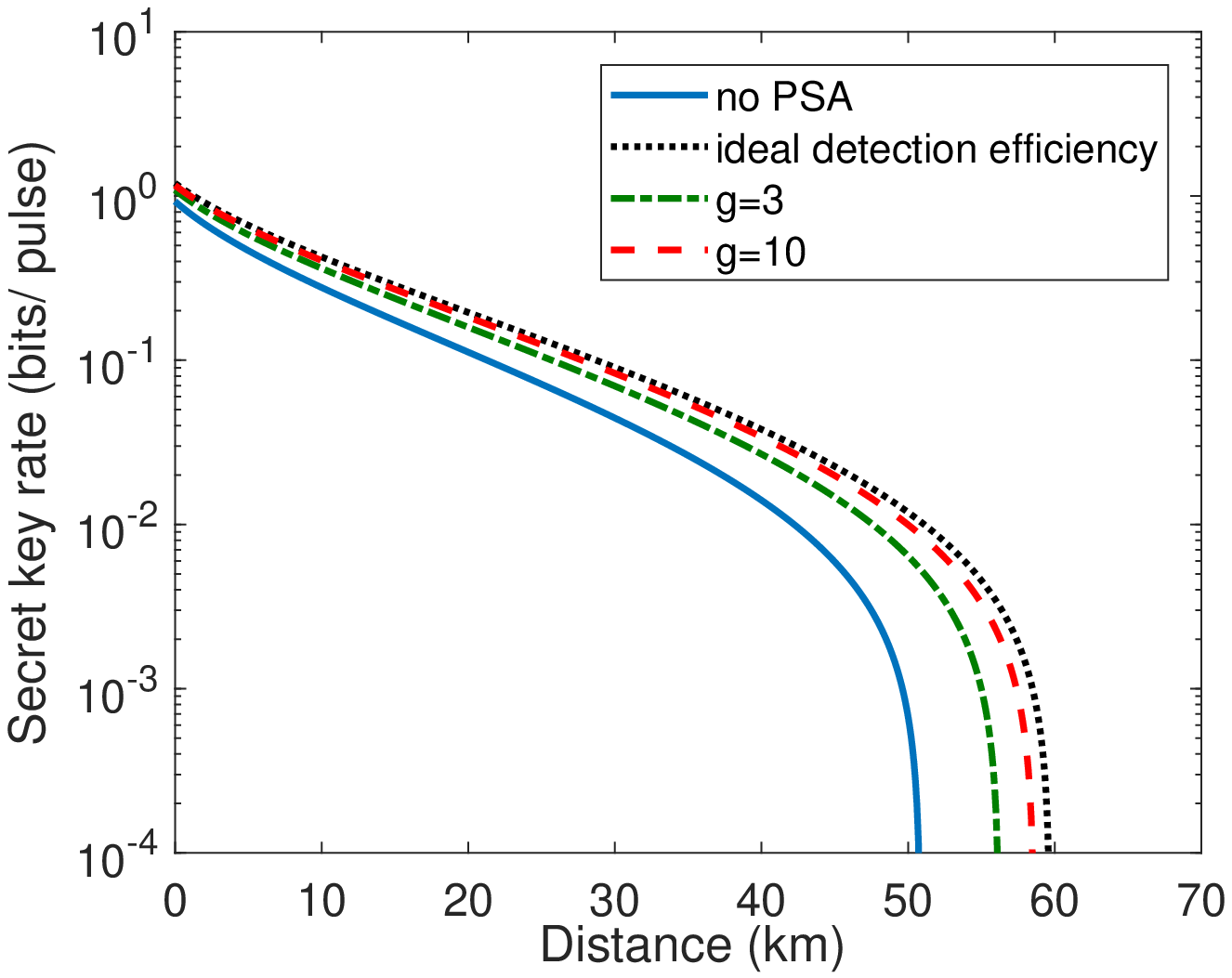}}
\caption{The secret key rate as a function of the transmittion distance. Simulation parameters are: channel excess noise ${{\varepsilon _c} = 0.01}$, the variance of the initial EPR state ${V=40}$, electronic noise ${\eta_e=0.9}$, detection efficiency ${\eta_d=0.6}$ except for the black curve (${\eta_d=1}$), and reconciliation efficiency ${\beta  = 0.956}$~\cite{zhou2019continuous}.
}
%Green dash-dotted curve represents the secret key rate for ${g=3}$, red dashed curve represents the secret key rate for ${g=10}$. In order to make comparisons, blue solid curve and black dotted curve are shown as the secret key rate without the PSA and with ideal detection efficiency respectively.}
\end{figure}
Numerous simulation results are provided in this section to demonstrate the effect of applying the PSA to the modified practical homodyne model.
In figure 3, we show the secret key rate as a function of the transmission distance, the channel efficiency ${\alpha}$ is ${{\rm{0}}{\rm{.2}}\;{\rm{dB}}\cdot {\rm{k}}{{\rm{m}}^{ - 1}}}$ as standard optical fibers. The green dash-dotted curve is the secret key rate of gain ${g=3}$ of the PSA while the red dashed curve corresponds to the secret key rate of gain ${g=10}$ of the PSA. In order to make proper comparisons, we also display the results of the case where no PSA is considered in the system (blue solid curve) and the ideal case where the detection efficiency is set to 1 (black dotted curve). It can be seen that the secret key rate can be improved by the PSA at all transmission distance, and the higher the gain, the more improvement the secret key rate achieves. The secret key rate of ${g=10}$ is outperforming than that of ${g=3}$, and it is more approaching to the ideal detection efficiency case. When the gain of the PSA increases to infinitely high, the imperfection of the detection efficiency can be fully compensated.

\begin{figure}[htbp]
\centerline{\includegraphics[width=10cm]{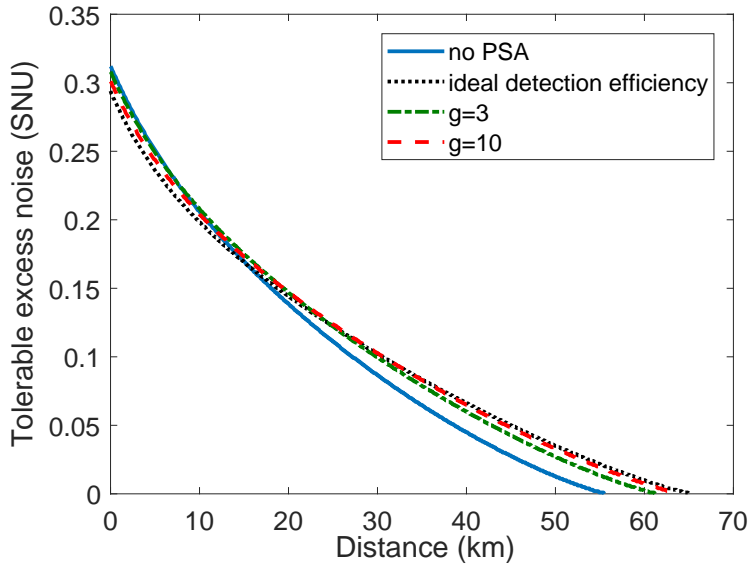}}
\caption{The maximal tolerable excess noise (SNU) as a function of the transmission distance. Simulation parameters are: channel excess noise ${{\varepsilon _c} = 0.01}$, the variance of the initial EPR state ${V=40}$, electronic noise ${\eta_e=0.9}$, detection efficiency ${\eta_d=0.6}$ except for the black curve (${\eta_d=1}$), and reconciliation efficiency ${\beta  = 0.956}$~\cite{zhou2019continuous}.}
%Green dash-dotted curve is tolerable excess noise for ${g=3}$, while red dashed curve is the result for ${g=10}$. Blue solid curve and black dotted curve are shown as the results without the PSA and with ideal detection efficiency respectively.
\end{figure}
In figure 4, we show the maximal tolerable excess noise as a function of the transmission distance, given different gain of the PSA ${g=3, 10}$, for the green dash-dotted curve and the red dashed curve respectively. The curve that corresponds to the case where no PSA is applied (blue solid curve) and the curve that corresponds to the ideal detection efficiency (black dotted curve) are also shown. The simulation results suggest that the improvement in the maximal tolerable excess noise is not so straightforward. As can be seen, when the transmission distance is less than 5km, the EB model without the assistance of the PSA can tolerate more excess noise, as the transmission distance increases, the PSA can help resist more excess noise, and the higher the gain, the more excess noise the model can tolerate.

%Next, we explain the principle of this model. Since the PSA is placed right before the BS that represents the detection efficiency, after the BS that represents the electronic noise, the PSA no longer amplifies the BS that imitates the electronic noise as well as its following output mode ${D}$, but only compensates the BS that corresponding to the detection efficiency and its outputs, thus in this model, the electronic noise can not be compensated by the PSA. However, for a infinitely high gain of ${g}$, the PSA can fully compensate the imperfections of the limited detection efficiency.

Finally, we show how the PSA affects the secret key rate with different modulation variance ${{V_A} = V - 1}$ in figure 5. We provide simulation results for three different transmission distance of ${30, 50, 80}$ km in green, blue and red curves respectively, and for different gains of ${g=3, 10}$ in dashed lines, dash-dotted lines, as well as the condition where no PSA is applied (solid lines) and the ideal condition where the detection efficiency is set to 1 (dotted lines). It can be seen from the simulation results, the PSA can be used to improve the secret key rates at all valid modulation variances. When the gain of the PSA reaches 10, the performance is very close to the ideal detection efficiency case. In the case where the transmission distance is 50km, we observe that the PSA can also increase the range of valid modulation variance.

\begin{figure}[tbp]
\centerline{\includegraphics[width=10cm]{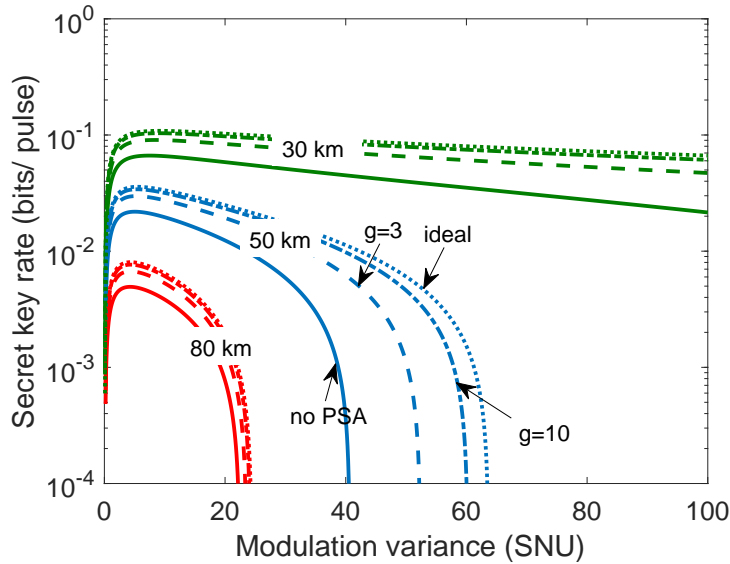}}
\caption{The secret key rate as a function of the modulation variance (SNU) ${{V_A} = V - 1}$, at different transmission distance. Simulation parameters are: channel excess noise ${{\varepsilon _c} = 0.01}$, the variance of the initial EPR state ${V=40}$, electronic noise ${\eta_e=0.9}$, detection efficiency ${\eta_d=0.6}$ except for the black curves (${\eta_d=1}$), and reconciliation efficiency ${\beta  = 0.956}$~\cite{zhou2019continuous}.}
%Red curves are secret key rates at 80km, blue curves and green curves are corresponding to the trancemission distance of 50km and 30km respectively. The gain of PSA is set as 3, 10 for the dashed lines and dash-dotted lines, solid lines are results when no PSA is applied, and dotted lines are results for ideal detection efficiency.
\end{figure}

\section{Conclusion}
In this paper, we demonstrate the modified practical homodyne detector model that can model the imperfections of the practical homodyne detector separately. It is proved to be essentially the same with the conventional practical homodyne detector model when the raw data from the practical implementations is normalized by the same shot-noise unit. The phase-sensitive amplifier is further applied to improve the performance of the modified practical homodyne detector model that can individually model the detection efficiency into trusted loss. Numerous simulation results suggest that the secret key rate can be improved by the phase-sensitive amplifier. When the gain of the phase-sensitive amplifier is high, the performance of the modified practical homodyne detector model can approach to the ideal detection efficiency case. Thus, we conclude that the phase-sensitive amplifier can be applied to improve the performance of the modified practical homodyne detector model.

\section*{Acknowledgements}
This work was supported by the Key Program of National Natural Science Foundation of China under Grant No. 61531003, the National Natural Science Foundation under Grant No. 61427813, the Fund of CETC under Grant No. 6141B08231115 and the Fund of State Key
Laboratory of Information Photonics and Optical Communications.

%\section*{References}
%\bibliographystyle{plain}
%\bibliography{OTC_PIA_bib}

\end{document}